\documentclass[twocolumn,showpacs,preprintnumbers,amsmath,amssymb]{revtex4}

\usepackage{amsmath}
\usepackage{graphicx}
\usepackage{calc}
\usepackage{bm}
\usepackage{float}

\begin{document}

\title{Metallic state in a strongly interacting spinless two-valley electron system\\ in two dimensions}
\author{M.~Yu.\ Melnikov}
\affiliation{Institute of Solid State Physics, Chernogolovka, Moscow District 142432, Russia}
\author{A.~A. Shashkin}
\affiliation{Institute of Solid State Physics, Chernogolovka, Moscow District 142432, Russia}
\author{V.~T. Dolgopolov}
\affiliation{Institute of Solid State Physics, Chernogolovka, Moscow District 142432, Russia}
\author{S.-H. Huang and C.~W. Liu}
\affiliation{Department of Electrical Engineering and Graduate Institute of Electronics Engineering, National Taiwan University, Taipei 106, Taiwan, and\\ National Nano Device Laboratories, Hsinchu 300, Taiwan}
\author{Amy Y.~X. Zhu and S.~V. Kravchenko}
\affiliation{Physics Department, Northeastern University, Boston, Massachusetts 02115, USA}

\begin{abstract}
We have studied the strongly interacting, two-valley two-dimensional (2D) electron system in ultrahigh mobility SiGe/Si/SiGe quantum wells in parallel magnetic fields strong enough to completely polarize the electron spins thus making the electron system ``spinless''. It occurs that the metallic temperature dependence of the resistivity, although weaker than that in the absence of magnetic field, still remains strong even when the spin degree of freedom is removed. Several independent methods have been used to establish the existence of the genuine MIT in the spinless two-valley 2D system. This is in contrast to the previous results obtained on more disordered silicon samples, where the polarizing magnetic field causes a complete quench of the metallic temperature behavior.
\end{abstract}
\maketitle

Spin and valley degrees of freedom in 2D electron systems have recently attracted much attention due to rapidly developing fields of spintronics and valleytronics (see, \textit{e.g.}, Refs.~\cite{behnia2012polarized,zhu2012field,schaibley2016valleytronics,zhu2017emptying}).  The existence of the zero-magnetic-field metallic state and the metal-insulator transition (MIT) in strongly interacting 2D electron systems is intimately related to the existence of these degrees of freedom \cite{lee1985disordered,punnoose2001dilute,punnoose2005metal,fleury2008many}.  The MIT in two dimensions was theoretically envisioned based on renormalization group analysis (see Ref.~\cite{lee1985disordered} for a review), and indeed, strong metallic temperature dependence of the resistivity $\rho(T)$ at temperatures low compared to the Fermi energy was experimentally observed in a variety of 2D systems \cite{abrahams2001metallic,kravchenko2004metal,shashkin2005metal,pudalov2006metal,spivak2010transport,qiu2018new,shashkin2019recent}.  It is worth noting that the metallic ($d\rho/dT>0$) temperature dependence of the resistivity at sub-Kelvin temperatures (especially, a weak metallic $\rho(T)$ dependence like, \textit{e.g.}, that of Ref.~\cite{gunawan2007spin}) does not by itself mean that the system is in the metallic state at zero temperature and, therefore, additional criteria have been used to verify the existence of the MIT \cite{shashkin2001metal}. Evidence that the MIT in clean 2D electron systems is driven by interactions is supplied by the observation of a strongly enhanced effective electron mass near the transition \cite{shashkin2002sharp,mokashi2012critical,melnikov2017indication}. The importance of the strong interactions in 2D electron systems has been confirmed recently in the observation of the formation of a quantum electron solid in silicon metal-oxide-semiconductor field-effect transistors (MOSFETs) \cite{brussarski2018transport}. The metallic state in single-valley 2D systems was predicted to be eliminated once the electron system becomes fully spin-polarized by a magnetic field parallel to the 2D plane \cite{lee1985disordered}. On the other hand, the electron spectrum in silicon-based 2D systems contains two almost degenerate valleys, which should further promote the metallicity \cite{punnoose2001dilute,punnoose2005metal,fleury2008many}. Therefore, the metallic state can, in principle, survive in these systems in the presence of spin-polarizing magnetic fields.  However, in disordered 2D systems, the irregularities of the interface lead to a finite intervalley scattering rate $1/\tau_\perp$ that mixes the two valleys to effectively produce a single valley at low temperatures \cite{punnoose2010renormalization1,punnoose2010test}, leading to an insulating state in spin-polarizing magnetic fields. The suppression of both the metallic regime and the metallic temperature behavior of the resistance was experimentally observed in Si MOSFETs \cite{dolgopolov1992properties,simonian1997magnetic,pudalov1997instability,shashkin2001metal,li2017resistivity}, $p$-type GaAs/AlGaAs heterostructures \cite{yoon2000parallel,gao2006spin} and narrow AlAs quantum wells \cite{vakili2004spin} placed in a parallel magnetic field, $B_\parallel\geq B^*$, strong enough to completely polarize the electron spins. It is important to note that carriers in both latter systems occupy only a single valley. Alternative concepts of the MIT like Wigner-Mott transition \cite{camjayi2008coulomb}, percolation transition \cite{tracy2009observation,dassarma2013two}, and liquid-solid transition \cite{spivak2003phase} do not explicitly discuss the spin and valley effects.

Here we report studies of the temperature dependence of the resistivity of the strongly interacting, spin-polarized (or ``spinless'') two-valley 2D electron system in ultrahigh mobility SiGe/Si/SiGe quantum wells.  We find that the metallic temperature behavior of the resistivity, $\rho(T)$, although weaker than that in zero magnetic field \cite{melnikov2019quantum}, remains strong in fully spin-polarizing parallel magnetic fields, in contrast to that in the best Si MOSFETs \cite{shashkin2001metal}.  Two independent methods (sign change of $d\rho/dT$ and vanishing activation energy and nonlinearity of current-voltage characteristics on the insulating side) yield critical electron densities for the MIT that coincide within the experimental uncertainty thus confirming the existence of the MIT in a spinless two-valley electron system.  The fact that the spinless electrons behave differently as compared to those in Si MOSFETs can be attributed to different intervalley scattering rates: the level of short-range disorder potential in our samples is some two orders of magnitude lower than that in the least disordered Si MOSFETs, hence the intervalley scattering rate should be small compared to that in Si MOSFETs, corresponding to the case of two distinct valleys.  The critical electron density, $n_{\text c}(B^*)$, for the MIT in the spinless electron system is higher by a factor of about 1.2 compared to $n_{\text c}(0)$ in zero magnetic field, which is consistent with theoretical calculations \cite{dolgopolov2017spin}.  The observed metallic temperature behavior is comparable to that in strongly interacting, spin-unpolarized single-valley 2D systems in the cleanest $p$-type GaAs/AlGaAs heterostructures, which indicates the same role of spins and distinct valleys with respect to the existence of the metallic state and the MIT.

\begin{figure}
\scalebox{0.59}{\includegraphics[angle=0]{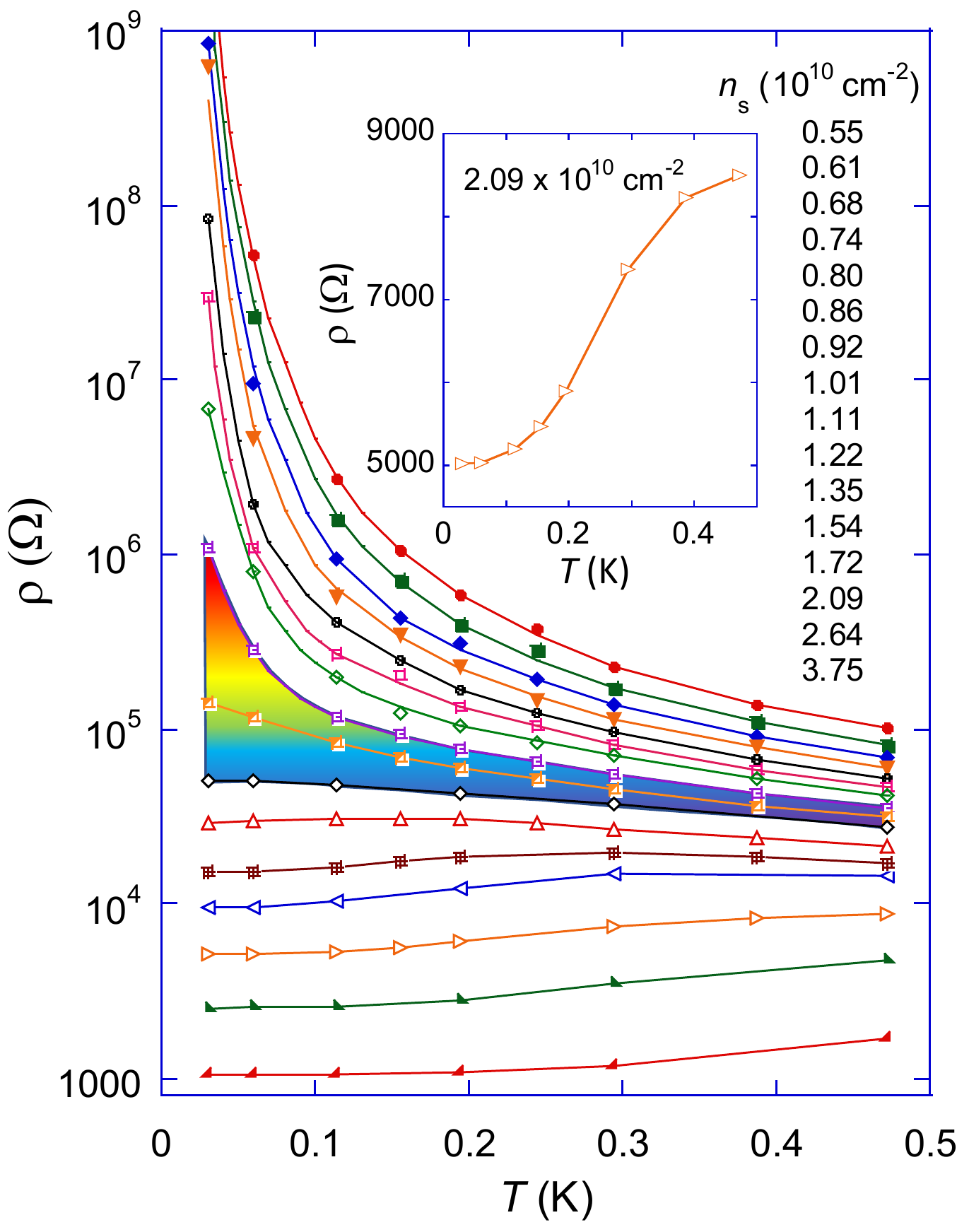}}
\caption{\label{fig1} Resistivity of an electron system in a SiGe/Si/SiGe quantum well placed in the spin-polarizing magnetic field $B^*$ as a function of temperature for different electron densities.  The magnetic fields used are spanned in the range between approximately 1 and 2~T.  The critical region around the MIT is color gradated.  The inset shows a closeup view of $\rho(T)$ for $n_{\text s}=2.09\times10^{10}$~cm$^{-2}$.}
\end{figure}

The samples studied are SiGe/Si/SiGe quantum wells of unprecedented quality similar to those described in detail in Refs.~\cite{melnikov2015ultra,melnikov2017unusual}. The peak electron mobility in these samples reaches 240~m$^2$/Vs. The approximately 15~nm wide silicon (001) quantum well is sandwiched between Si$_{0.8}$Ge$_{0.2}$ potential barriers. The samples were patterned in Hall-bar shapes with the distance between the potential probes of 150~$\mu$m and width of 50~$\mu$m using standard photo-lithography. Measurements were carried out in an Oxford TLM-400 dilution refrigerator. Data on the metallic side of the transition were taken by a standard four-terminal lock-in technique in a frequency range 0.5--11~Hz in the linear regime of response. On the insulating side of the transition, the resistance was measured with {\it dc} using a high input impedance electrometer. Since in this regime, the current-voltage ($I$-$V$) curves are strongly nonlinear, the resistivity was determined from ${\rm d}V/{\rm d}I$ in the linear interval of $I$-$V$ curves, as $I\rightarrow0$.

\begin{figure}[b]%\vspace{2mm}
\scalebox{0.48}{\includegraphics[angle=0]{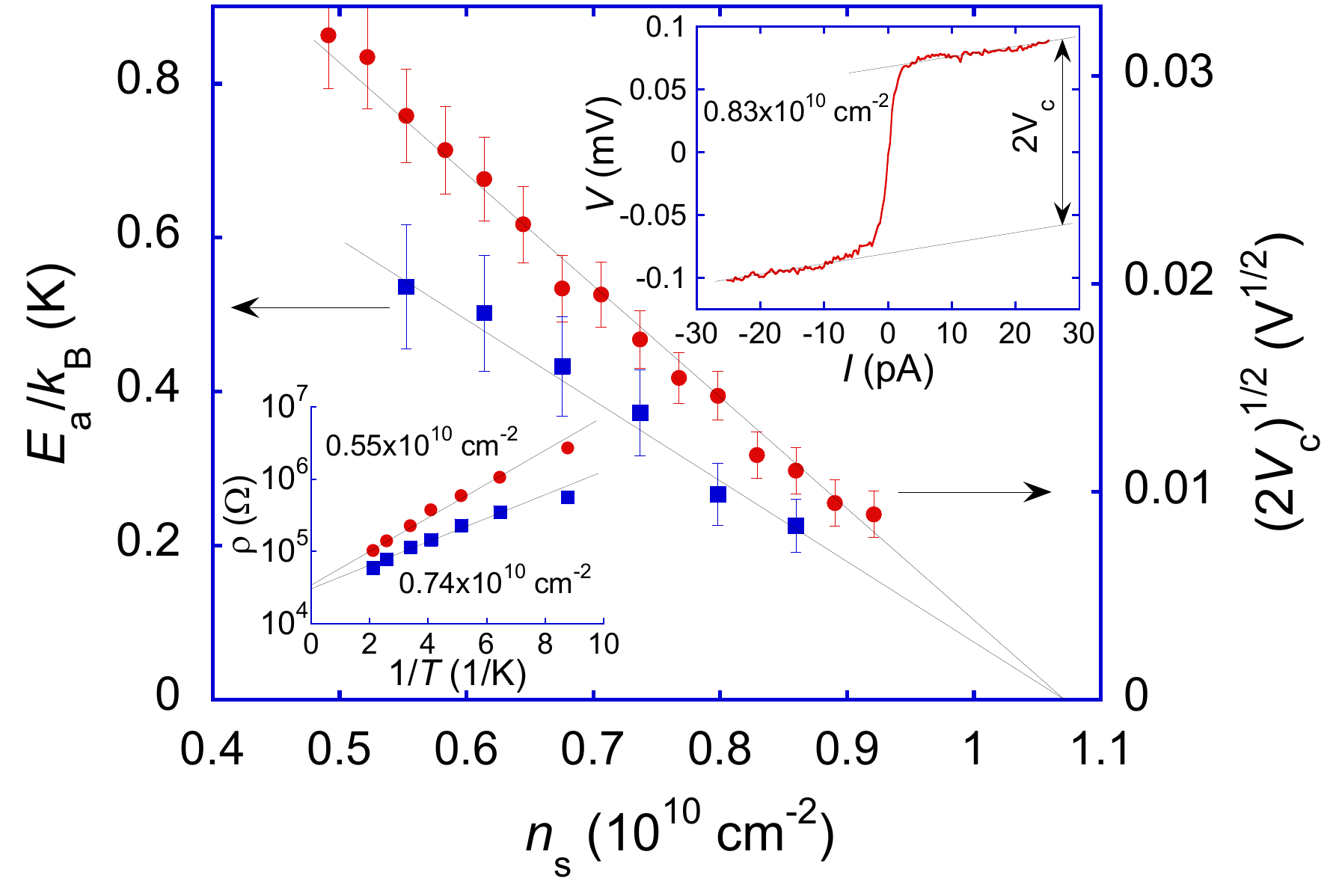}}
\caption{\label{fig2} Main panel: activation energy, $E_{\text a}$, and square root of the threshold voltage, $V_{\text c}^{1/2}$, \textit{vs}.\ electron density.  Solid lines correspond to the best linear fits.  Upper inset: a typical $I$-$V$ dependence on the insulating side of the MIT at $T=30$~mK.  Lower inset: Arrhenius plots of the temperature dependence of the resistivity for two electron densities on the insulating side.}
\end{figure}

In Fig.~\ref{fig1} we plot the resistivity $\rho(T)$, measured in a parallel magnetic field strong enough to polarize the electron spins, for 16 different electron densities $n_{\text s}$.  The magnetic field for complete spin polarization is density-dependent and was determined by the saturation of the $\rho(B_\parallel)$ dependence, which corresponds to the lifting of the spin degeneracy \cite{okamoto1999spin,vitkalov2000small}; the magnetic fields used in our experiments fell within the range between approximately 1 and 2~T. At the lowest temperatures, the resistivity exhibits a strong metallic temperature dependence ($d\rho/dT>0$) for electron densities above a certain critical value, $n_{\text c}(B^*)$, and an insulating behavior ($d\rho/dT<0$ with resistivity diverging as $T\rightarrow0$) for lower densities.  The critical region between the insulating and metallic behaviors is color-gradated.  Assuming that the extrapolation of $\rho(T)$ to $T=0$ is valid and taking into account that the curve separating metallic and insulating regimes should be tilted \cite{punnoose2005metal}, we identify the critical density for the metal-insulator transition $n_{\text c}(B^*)=(1.11\pm0.05)\times10^{10}$~cm$^{-2}$ in a way similar to the case of $B=0$ \cite{melnikov2019quantum}. The $\rho(T)$ dependences on the metallic side of the transition at $n_{\text s}$ just above the critical density are non-monotonic: while at temperatures exceeding a density-dependent value $T_{\text m}$, the derivative $d\rho/dT$ is negative (``insulating-like''), it changes sign at temperatures below $T_{\text m}$. The measurements were restricted to 0.5~K that is the highest temperature at which the saturation of the $\rho(B_\parallel)$ dependence could still be achieved; the restriction is likely to reflect the degeneracy condition for the dilute electron system with low Fermi energy. It is worth noting that the observed metallic temperature dependence of the resistivity indicates the presence of electron backscattering and of a short-range disorder potential (see discussions in Ref.~\cite{melnikov2015ultra}).

On the metallic side of the transition ($n_{\text s}>n_{\text c}(B^*)$), the maximum resistivity drop with decreasing temperature below 0.5~K almost reaches a factor of 2 (see the line plot in the inset in Fig.~\ref{fig1}), which is weaker compared to more than an order-of-magnitude drop in this system at $B=0$ \cite{melnikov2019quantum}.  Still, the metallic temperature behavior of spinless electrons in SiGe/Si/SiGe quantum wells remains strong and similar to that observed in $p$-type GaAs/AlGaAs heterostructures in zero magnetic field \cite{hanein1998the,gao2006spin}.

The critical density for the MIT can also be inferred from two additional criteria not requiring the extrapolation of the data to $T=0$: namely, vanishing of the activation energy and nonlinearity of the current-voltage characteristics on the insulating side of the transition ($n_{\text s}<n_{\text c}(B^*)$) \cite{shashkin1994insulating}.  The temperature dependences of the resistivity have an activation character on the insulating side in the vicinity of the transition (see the lower inset in Fig.~\ref{fig2}); the activation energy $E_{\text a}$ is plotted in the main panel of Fig.~\ref{fig2} as a function of $n_{\text s}$.  The dependence is linear, which corresponds to the constant thermodynamic density of states near the critical point, and extrapolates to zero at $n_{\text c}(B^*)=(1.07\pm0.03)\times10^{10}$~cm$^{-2}$ which coincides, within the experimental uncertainty, with the value of $n_{\text c}(B^*)$ determined from the temperature derivative criterion.

\begin{figure}
\scalebox{0.5}{\includegraphics[angle=0]{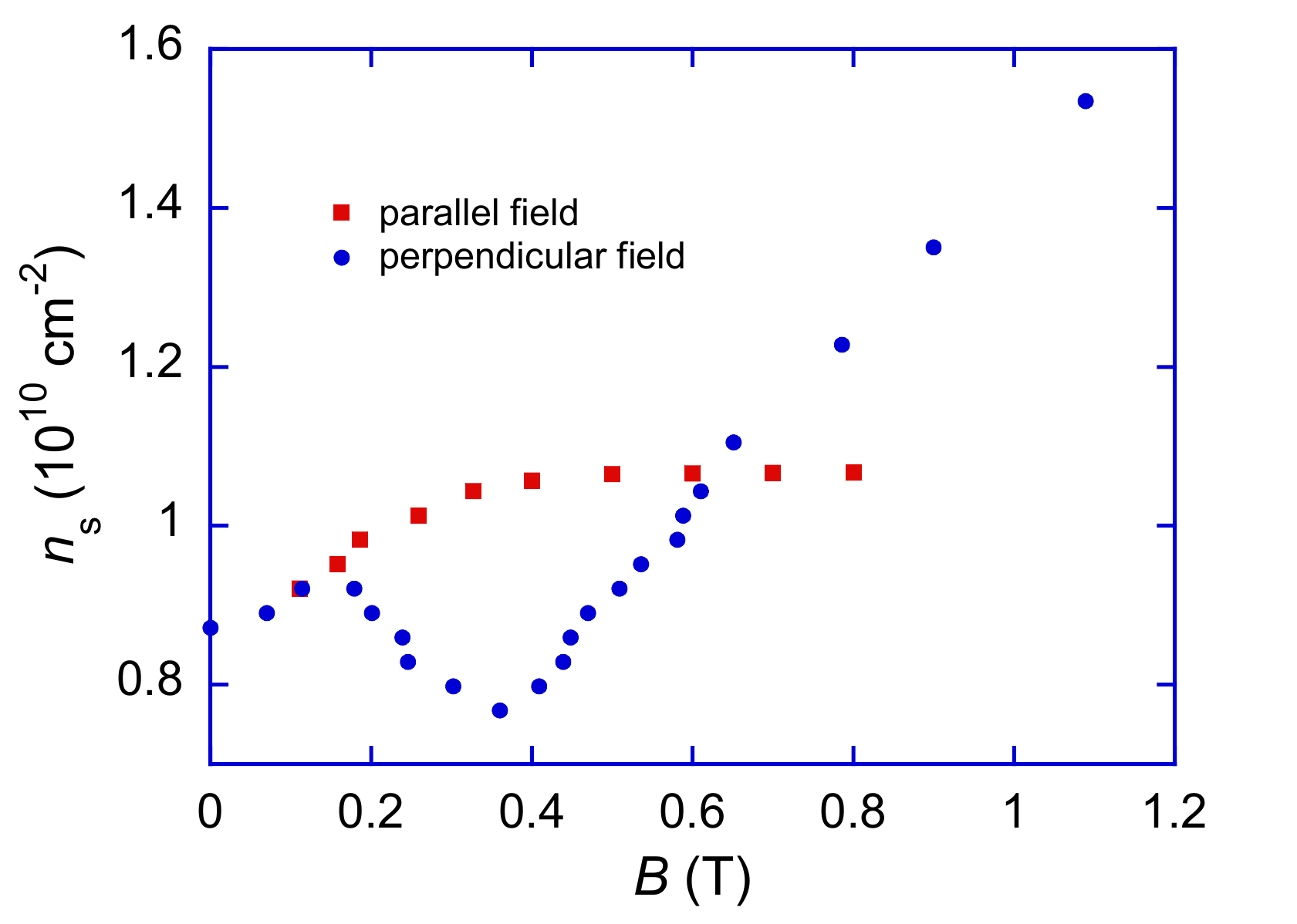}}
\caption{\label{fig3} Critical electron density for the MIT as a function of the magnetic field parallel (squares) and perpendicular (circles) to the 2D plane.  The cutoff resistivity for the MIT was chosen at $\rho=200$~k$\Omega$ at a temperature of 30~mK (see text).}
\end{figure}

A typical $I$-$V$ curve measured on the insulating side of the MIT ($n_{\text s}<n_{\text c}(B^*)$) is shown in the upper inset to Fig.~\ref{fig2}.  The $V(I)$ dependence obeys Ohm's law in a very narrow interval of currents $\left|I\right|\lesssim1$~pA and almost saturates at higher excitation currents.  Such a threshold behavior of the $I$-$V$ curves has been explained within the concept of the breakdown in the insulating phase \cite{polyakov1993conductivity,shashkin1994insulating} that occurs when the localized electrons at the Fermi level gain enough energy to reach the mobility edge in an electric field $V_{\text c}/d$ over a distance given by the localization length, $L$: $e V_{\text c}L/d = E_{\text c}-E_{\text F}$, where $d$ is the distance between the potential probes.  The activation energy and the threshold voltage are related through the localization length, which is temperature-independent and diverges near the transition as $L(E_{\text F})\propto(E_{\text c}-E_{\text F})^{-s}$ with exponent $s$ close to unity \cite{shashkin1994insulating}. This corresponds to a linear dependence $V_{\text c}^{1/2}(n_{\text s})$ near the MIT, as seen in the main panel of Fig.~\ref{fig2}.  The dependence extrapolates to zero at the same electron density as the $E_{\text a}(n_{\text s})$ dependence.

In Fig.~\ref{fig3} we plot the critical density for the MIT as a function of the magnetic field parallel ($B_{\parallel}$, squares) and perpendicular ($B_\perp$, circles) to the 2D plane.  To determine the critical densities, we have chosen the cutoff resistivity for the MIT at $\rho=200$~k$\Omega$, which is close to the value of the critical resistivity for the zero-field MIT at the lowest accessible temperatures (note that the behavior of the $n_{\text{c}}(B)$ phase diagram is only weakly sensitive to the particular cutoff value). In both cases, the critical density initially monotonically increases with the magnetic field; the data for the parallel and perpendicular magnetic fields coincide up to $B\approx0.15$~T.  In higher parallel magnetic fields, the critical density continues to monotonically increase before eventually saturating at $B_\parallel\gtrsim0.5$~T at the level roughly a factor of 1.2 higher than that in zero field.  This corresponds to fully polarized electron spins.  In perpendicular field, the $n_{\text{c}}(B_\perp)$ dependence is non-monotonic.  It has a minimum corresponding to the Landau level filling factor $\nu=n_{\text{s}}hc/eB_\perp=1$, which is similar to the re-entrant behavior of the insulating phase that was observed earlier in Si MOSFETs and GaAs/AlGaAs heterostructures \cite{diorio1990magnetic,dolgopolov1992properties,jiang1993observation,kravchenko1995global,qiu2012connecting}. Then, $n_{\text{c}}(B_\perp)$ monotonically increases with $B_\perp$ so that its slope reaches $\nu\approx0.3$ in the high-field limit. This is in contrast to the slope of $\nu\approx0.5$ observed in more disordered 2D electron system in Si MOSFETs in the extreme quantum limit \cite{dolgopolov1992metal} and interpreted as an indication of a single-particle localization of electrons below half-filling of the lowest Landau level (for more on this, see Ref.~\cite{melnikov2019quantum}).

For spin-unpolarized electrons in zero magnetic field, in both Si MOSFETs and SiGe/Si/SiGe quantum wells, the derivative sign-change criterion and the temperature-independent criterion based on vanishing activation energy and nonlinearity of the $I$-$V$ curves yield the same critical electron density for the MIT: $n_\text{c}(0)\approx8.0\times10^{10}$~cm$^{-2}$ in Si MOSFETs \cite{shashkin2001metal} and $n_\text{c}(0)\approx0.87\times10^{10}$~cm$^{-2}$ in SiGe/Si/SiGe quantum wells \cite{melnikov2019quantum}.  However, spinless electrons behave differently in the two systems, which can be attributed to different intervalley scattering rates.  In Si MOSFETs, where the short-range disorder level is two orders of magnitude higher than that in the samples studied here, the strong intervalley scattering mixes two valleys at low temperatures effectively producing a single valley, and the derivative criterion does not at all yield a critical density for spinless electrons. Since the second criterion mentioned above holds, this leaves uncertain the existence of a metal-insulator transition in this system \cite{shashkin2001metal}.  In contrast, in ultrahigh mobility SiGe/Si/SiGe quantum wells, the metallic temperature dependence of the resistivity remains strong, and both the above-mentioned criteria yield critical densities that coincide within the experimental uncertainty confirming the existence of the MIT in this 2D system of spinless electrons that retain another, valley degree of freedom.  The observed metallic temperature behavior is comparable to that in strongly interacting, spin-unpolarized single-valley 2D systems in the cleanest $p$-type GaAs/AlGaAs heterostructures, which shows that the role of distinct valleys in the electron spectrum is equivalent to the role of spins in regard to the existence of the metallic state and the MIT in two dimensions.

It is interesting to compare the ratio $n_{\text c}(B^*)/n_{\text c}(0)$ of the measured critical densities for the MIT to that calculated in Ref.~\cite{dolgopolov2017spin}.  According to the calculations, the increase of the critical electron density for the Anderson transition in a strongly interacting 2D electron system with increasing $B_\parallel$ is due to the exchange and correlation effects, and the ratio between the critical electron densities for fully spin-polarized and unpolarized electron systems is independent of the density of impurities and is equal to $\approx1.33$. This value is consistent with our experimental results. Noticeably, a similar, although somewhat stronger, suppression of the metallic regime was previously observed in Si MOSFETs where the localization of fully spin-polarized electrons occurs at the electron density by a factor of about 1.4 higher compared to the localization of unpolarized electrons in zero magnetic field \cite{dolgopolov1992properties,shashkin2001metal,eng2002effects}.

In summary, we have experimentally shown that the strongly interacting, spinless electron system in ultrahigh mobility SiGe/Si/SiGe quantum wells exhibits a well-defined MIT that we attribute to the existence of two distinct valleys in its spectrum.  The drop of the resistivity with decreasing temperature on the metallic side of the metal-insulator transition in the spinless system is weaker compared to that in $B=0$ but is comparable to the drop in spin-unpolarized single-valley 2D systems in the cleanest $p$-type GaAs/AlGaAs heterostructures.  This shows that in ultra-clean strongly interacting 2D systems, the valleys play the same role as spins in regard to the existence of the metallic state and the MIT.

We gratefully acknowledge discussions with D. Heiman. The ISSP group was supported by RFBR Grants No.\ 18-02-00368 and No.\ 19-02-00196, RAS, and the Russian Government Contract.  The NTU group acknowledges support by the Ministry of Science and Technology, Taiwan (Project No.\ 107-2218-E-002-044, 107-2622-8-002-018, and 108-3017-F-009-003) and Ministry of Education, Taiwan (project No.\ NTU-CC-108L891701). The Northeastern group was supported by NSF Grant No.\ 1309337 and BSF Grant No.\ 2012210.

%\bibliography{references}

%\bibliographystyle{apsrev4-1}
%\bibliographystyle{apsrmp}
%\bibliographystyle{apsrev}
%\bibliographystyle{naturemag}

\end{document}